# Deep Learning Improves Dataset Recovery for High Frame Rate Synthetic Transmit Aperture Imaging

Jingke Zhang, *Graduate Student Member, IEEE*, and Jianwen Luo*, *Senior Member, IEEE*

*Abstract*—**Synthetic transmit aperture (STA) imaging benefits from the two-way dynamic focusing to achieve optimal lateral resolution and contrast resolution in the full field of view, at the cost of low frame rate (FR) and low signal-to-noise ratio (SNR). In our previous studies, compressed sensing based synthetic transmit aperture (CS-STA) and minimal $l_2$-norm least squares (LS-STA) methods were proposed to recover the complete STA dataset from fewer Hadamard-encoded plane wave (PW) transmissions. Results demonstrated that, compared with STA imaging, CS/LS-STA can maintain the high resolution of STA in the full field of view and improve the contrast in the deep region with increased FR. However, these methods would introduce errors to the recovered STA datasets and subsequently produce severe artifacts to the beamformed images, especially in the shallow region. Recently, we discovered that the theoretical explanation for the error introduced in the LS-STA-based recovery is that LS-STA method neglects the null space component of the real STA data. To deal with this problem, we propose to train a convolutional neural network (CNN) under the null space learning framework (to estimate the missing null space component) for high-accuracy recovery of the STA dataset from fewer Hadamard-encoded PW transmissions. The mapping between the low-quality STA dataset (recovered using the LS-STA method) and the corresponding high-quality STA dataset (obtained using full Hadamard-encoded STA imaging, HE-STA) was learned by the network from phantom and *in vivo* samples. The performance of the proposed CNN-STA method was compared with the baseline LS-STA, conventional STA, and HE-STA methods, in terms of visual quality, normalized root-mean-square error (NRMSE), generalized contrast-to-noise ratio (gCNR), and lateral full width at half maximum (FWHM). The results demonstrate that the proposed method can greatly improve the recovery accuracy of the STA datasets (lower NRMSE) and therefore effectively suppress the artifacts presented in the images (especially in the shallow region) obtained using the LS-STA method (with a gCNR improvement of 0.4 in the cross-sectional carotid artery images). In addition, the proposed method can maintain the high lateral resolution of STA with fewer (as low as 16) PW transmissions, as LS-STA does.**

This work was supported in part by the National Natural Science Foundation of China (61871251 and 62027901), Beijing Natural Science Foundation (No. M22018) and Tsinghua University Spring Breeze Fund (No. 2021Z99CFY025)

Jingke Zhang and Jianwen Luo* are with the Department of Biomedical Engineering, School of Medicine, Tsinghua University, Beijing 100084, China (e-mail: luo_jianwen@tsinghua.edu.cn)

## I. INTRODUCTION

Thanks to the two-way dynamic focusing, synthetic transmit aperture (STA) imaging can achieve the highest spatial (lateral) resolution and contrast resolution among all the imaging sequences [1]. In STA imaging, all the $N$ transducer elements are activated individually and sequentially to transmit spherical waves and to obtain $N$ low-quality images. Thereafter, transmit focusing can be resynthesized at all image points by coherently summing these $N$ images, to obtain the final high-resolution STA image. However, owing to the limited energy transmitted by the individual element in each transmission, STA imaging severely suffers from the low signal-to-noise ratio (SNR) issue (especially in the deep region). In addition, as STA imaging requires activating all the elements sequentially, its frame rate is limited by a large number of transmissions.

Several methods have been proposed to improve the SNR and frame rate of STA imaging. Multi-element STA method uses sub-apertures to transmit diverging waves to increase the transmitted energy and reduce the number of transmissions [2]. However, the reduced angular spread of the diverging transmit beam would affect the effect of transmit focusing. By considering the transmit focal points as virtual sources of diverging waves, which propagate toward and away from the transducer during beamforming, high-SNR STA imaging can be achieved with focused transmissions [3]. Although these methods can improve the SNR and the frame rate of STA imaging at the same time, they cannot really acquire the complete STA dataset, which contains the entire backscattered echoes corresponding to all transmit-receive pairs. In other words, these methods cannot achieve optimal transmit focusing.

Coding methods aim at recovering the individual transmit element responses on receive after high-energy sub-aperture or full-aperture transmissions. Hadamard coding [4] and S-sequence [5] utilize the orthogonal and invertible properties of the spatial codes (applied across the aperture) to accurately decode the high-SNR complete STA dataset from the encoded transmissions. The elements across the aperture can also be temporally encoded by applying orthogonal transmit delay [6]. By transmitting broadband and temporally extended coded pulse, the energy transmitted by each element can be increased to improve the SNR of the received echoes [7]. Although these methods can recover the complete STA dataset, they cannot reduce the number of required transmissions to form a single image. In summary, the conventional methods typically have to find a balance between the frame rate and ideal focusing.

In recent years, researchers have attempted to increase the frame rate and SNR of STA imaging (at the same time) by recovering the complete STA dataset from fewer high-energy transmissions. In the frequency domain, Bottenus established a linear relationship between the complete STA dataset and the focused wave dataset, to reconstruct STA images with greatly improved SNR [8]. However, the increase in frame rate brought by this method is limited [8].

In the time domain, Liu et al proposed the compressed sensing (CS)-based STA method (CS-STA) to recover the complete STA dataset from fewer encoded plane wave (PW) transmissions, using the CS technique in linear array configuration [9]. Results demonstrated that the CS-STA method can maintain the high spatial resolution of STA imaging with only a quarter of (or less) transmissions, while achieving greatly improved contrast and SNR (especially in the deep region). Subsequently, the CS-STA method has been validated in phased array [10], convex array [11], and matrix array [12] configurations. Thereafter, the CS-STA method has been extended from the radio-frequency (RF) domain to the in-phase/quadrature (IQ) domain for acceleration and adaptation to clinical ultrasound systems (which typically record baseband IQ signals) [13]. Moreover, Zhang et al proposed to substitute the time-consuming iterative CS method with the minimal $l_2$-norm least squares method (LS-STA), to significantly reduce the complete STA dataset recovery time from tens of minutes (for CS-STA) to hundreds of milliseconds [14]. However, both CS and LS techniques would introduce errors to the recovered complete STA dataset.

The quality of images obtained using the CS/LS-STA methods is affected comprehensively by the recovery error and SNR of the recovered STA dataset. For example, if the imaging target is located in shallow regions, from which the STA transmissions could acquire backscattered echoes with an acceptable SNR, the further increase in SNR brought by the encoded PW transmissions would not be of great help for improving the image quality. Therefore, the recovery error would be the dominant factor and lead to a decrease in image quality. In contrast, if the STA acquisitions suffer from severe noise, the encoded transmissions of CS/LS-STA could greatly increase the SNR of received echoes to improve the image quality, even in the presence of recovery error. In a simple word, the CS/LS-STA methods typically perform better than STA imaging in the deep region, but worse than STA imaging in the shallow region. The potential applications of the CS/LS-STA methods would be limited by their inconsistent performances in the depth direction. Therefore, it is important to reduce the recovery error for better image quality.

Deep learning has attracted great attention in the ultrasound imaging community, thanks to a great number of successful applications [15]. Deep learning was first introduced to learn a non-linear mapping from the low-quality image acquired with 3 PWs to the high-quality image acquired with 31 PWs [16]. In this way, the imaging frame rate can be increased without sacrificing the image quality. Ever since, many methods have been proposed to further improve the performance of this strategy [17][18][19][20][21][22][23]. Deep learning can also be incorporated into the ultrasound beamforming procedure. For example, it was utilized to efficiently calculate the optimal receive apodizations [24] or short-lag spatial coherence (SLSC) [25], which are typically calculated using the conventional time-consuming adaptive beamforming methods [26]. The above methods require the delayed RF data as the input to the network. Some researchers attempted to skip the delay step and directly feed the network with the raw RF data for beamforming [27][28][29]. The last type of methods was proposed to pre-process the RF channel data with deep learning to improve the performance of conventional beamformers. Specifically, deep learning was used to suppress the reverberation noise [30] and off-axis scattering noise [31], or to recover the complete RF data from subsampled RF data [32][33]. As the deep learning techniques achieve very good performances in ultrasound beamforming tasks, including the pre-processing of the RF data, in this study, we propose to introduce the deep learning technique into the recovery of the STA dataset to reduce the error and to improve the subsequent image quality.

The rest of this paper is organized as follows. Section II presents the theoretical basis of recovering the STA dataset from the encoded PW transmissions, and reviews the principles of the CS-STA and LS-STA methods. Section III introduces the detailed processing steps of the proposed method, the experimental setups, the training and testing datasets, and the evaluation metrics. Section IV presents the results. Sections V and VI discuss and conclude this work, respectively.

## II. RELATED WORKS

### A. Theory Basis

According to the superposition principle of the linear acoustic theory, the overall response caused by the multi-element transmission is the sum of the responses that would have been caused by each transducer element individually. That is, the backscattered echo received after apodized PW transmission is the linear combination of the received echoes of all the STA transmissions, and the linear combination coefficients are the apodizations applied in the PW transmission [9].

Assume that $x_{n,j}(t)$ denotes the RF data recorded at time $t$ by the $j$th ($1 \le j \le N$) element when only the $n$th element is activated for transmission, and $y_{m,j}(t)$ denotes the RF data recorded at time $t$ by the $j$th ($1 \le j \le N$) element when the $m$th apodized PW is transmitted. A linear measurement model can be established as:

$$y_{m,j}(t) = \sum_{n=1}^{N} H_{m,n} x_{n,j}(t) \tag{1}$$

where $H_{m,n}$ denotes the apodization applied on the $n$th element in the $m$th PW transmission.

By omitting the time symbol $t$ and vectorizing Eq. (1), the measurement model can be generalized as:

$$\boldsymbol{y}_j = \boldsymbol{H}\boldsymbol{x}_j \qquad (2)$$

where $\boldsymbol{y}_j = [y_{1,j},\ y_{2,j}, \ldots, y_{M,j}]^T$ and $\boldsymbol{x}_j = [x_{1,j},\ x_{2,j}, \ldots, x_{N,j}]^T$ denote the slow-time RF data (at time $t$) recorded by the $j^{th}$ element after $M$ times of PW transmissions and $N$ times of STA transmissions, respectively, and $\boldsymbol{H} \in \mathbb{R}^{M \times N}$ denotes the encoding (apodization) matrix.

As the linear measurement model has been established, recovery of the complete STA dataset from fewer encoded PW transmissions can be achieved by solving the corresponding inverse problem.

### B. Compressed Sensing Based Synthetic Transmit Aperture

CS-STA [9] assumes that the slow-time STA data $\mathbf{x}_j \in \mathbb{R}^N$ to be reconstructed can be sparsely represented in a sparse basis $\boldsymbol{\Psi}$ (such as a wavelet basis) as:

$$\boldsymbol{x}_j = \boldsymbol{\Psi}\boldsymbol{v}_j \qquad (3)$$

where most entries of $\boldsymbol{v}_j \in \mathbb{R}^n$ are zero or close to zero. Substituting Eq. (3) into Eq. (2), we can obtain:

$$\boldsymbol{y}_j = \boldsymbol{H}\boldsymbol{\Psi}\boldsymbol{v}_j \qquad (4)$$

Under the assumption that $\boldsymbol{v}_j$ is sparse, the CS technique can be used to solve the following optimization problem:

$$\hat{\boldsymbol{v}}_j = \underset{\boldsymbol{v}_j \in \mathbb{R}^n}{\arg\min} \|\boldsymbol{v}_j\|_1 \ subject\ to\ \|\boldsymbol{y}_j - \boldsymbol{H}\boldsymbol{\Psi}\boldsymbol{v}_j\|_2 \le \varepsilon \qquad (5)$$

where $\varepsilon$ denotes the tolerated error. Thereafter, the complete STA dataset can be obtained with Eq. (3).

Although previous results demonstrated that the CS-STA method can achieve satisfactory image quality, the assumption that the slow-time STA data are sparse in the transform domain is not strictly correct [9]. As a result, CS-based solving would inevitably introduce errors to the recovered STA dataset. In addition, owing to the iterative nature of the CS algorithm, the recovery time is very long (tens of minutes per frame).

### C. Minimal $l_2$-Norm Least-Squares Based Synthetic Transmit Aperture

LS-STA method [14] was proposed to overcome the time-consuming problem of the CS-STA-based recovery. Instead of assuming that $x_j$ is sparse in a transform domain, LS-STA assumes that $\mathbf{x}_j$ has the lowest energy and requires the encoding matrix $\boldsymbol{H}$ to be row full rank, i.e., rank($\boldsymbol{H}$)=$M$. Therefore, a minimal $l_2$-norm least squares solution can be obtained by solving the following problem:

$$\min\|x_j\|_2^2\ s.t.\ \boldsymbol{y}_j = \boldsymbol{H}\boldsymbol{x}_j \qquad (6)$$

By introducing Lagrange multipliers, the above problem has an analytical solution:

$$\hat{x}_j = \boldsymbol{H}^T(\boldsymbol{H}\boldsymbol{H}^T)^{-1}\boldsymbol{y}_j \qquad (7)$$

where $\boldsymbol{H}^T(\boldsymbol{H}\boldsymbol{H}^T)^{-1}$ is the right inverse $\boldsymbol{H}^\dagger$ of partial Hadamard matrix $\boldsymbol{H}$, which has linearly independent rows. In addition, thanks to the orthogonal property of $\boldsymbol{H}$, $\boldsymbol{H}\boldsymbol{H}^T$ equals a scaled identity matrix $N\boldsymbol{I}_M$. Therefore, a simple analytical solution can be obtained as:

$$\hat{x}_j = (\boldsymbol{H}^T\boldsymbol{y}_j)/N \qquad (8)$$

Previous results demonstrated that the LS-STA method is capable of achieving the same accuracy as the conventional CS-STA method for complete STA dataset recovery (when $\boldsymbol{H}$ is partial Hadamard matrix), but with a ~5,000 times faster recovery speed [14]. Therefore, LS-STA has a higher potential for real-time applications than CS-STA. However, the recovery accuracy and the subsequent image quality are still not optimal, especially for the anechoic/hypoechoic targets located in the shallow regions.

## III. METHODS

### A. Overview

The flow chart of the proposed convolutional neural network based synthetic transmit aperture imaging method (CNN-STA) is shown in Fig. 1. The network is fully convolutional and trained with null space learning, which will be introduced in detail in Sections III.C and III.B, respectively. The training inputs and labels for CNN-STA are the STA datasets recovered using the LS-STA method with partial-Hadamard-encoded and full-Hadamard-encoded PW transmissions, respectively. Note that when combined with full-Hadamard-encoded PW transmissions, LS-STA becomes the well-known Hadamard-encoded STA imaging (HE-STA) [4]. When using partial-Hadamard-encoded PW transmissions, the image quality (thanks to the high-energy PW transmissions) and the temporal resolution (thanks to the fewer transmissions) of STA imaging can be improved at the same time.

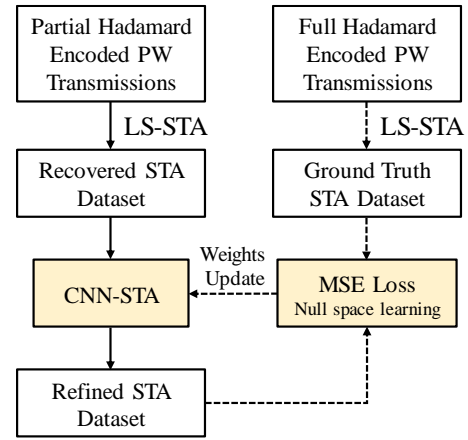

Fig. 1. Overview of the proposed CNN-STA method.

Let $\mathbf{Y} \in \mathbb{R}^{D \times N \times M}$ and $\mathbf{X} \in \mathbb{R}^{D \times N \times N}$ be three-dimensional (3D) cubes that represent the PW and STA raw RF data, where the three dimensions correspond to the samples (depth), receive channels (Rx), and transmissions (Tx), respectively, as shown in Fig. 2(a). Fig. 2(b) presents 2D (Rx-Tx) data corresponding to a given depth plane of the 3D cube. Fig. 2(c) presents a 1D (Tx) data corresponding to a given receive channel of the 2D plane. The CS-STA and LS-STA methods focus on recovering the 1D (Tx) slow-time data individually, which does not have a specific and clear feature. Therefore, it is difficult to provide accurate prior knowledge about the desired data during the recovery. In contrast, the 2D (Rx-Tx) STA data has at least two explicit and intuitive features, as shown in Fig. 2(b). First, according to the acoustic reciprocity, the backscattered echoes $x_{n,j}$ received by the $j^{th}$ element when the $n^{th}$ element is activated equals $x_{j,n}$. That is, the 2D (Rx-Tx) data are typically symmetrical. Second, the adjacent samples in the 2D (Rx-Tx) plane have close amplitudes (i.e., local smoothness) to form ripple-like structures, because the echoes backscattered to the adjacent elements have very close time-of-flights (TOF). Therefore, in this study, the recovered and ground truth 2D STA data are selected as the input-label pairs for the training of the network to exploit their explicit structural features.

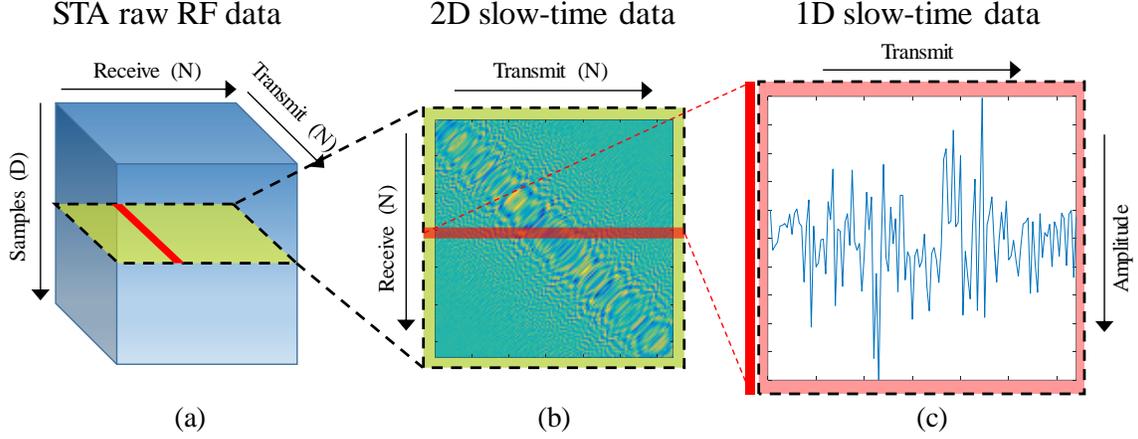

Fig. 2. Illustrations for (a) the complete 3D (samples-Rx-Tx) STA raw RF data, (b) 2D (Rx-Tx) STA data, and (c) 1D (Tx) slow-time STA data.

### B. Null Space Learning

Our goal is to accurately recover the STA dataset $\boldsymbol{x}'$ from fewer PW transmissions using a deep neural network. A natural idea is to train the network to learn a direct mapping from the acquired PW raw RF data $\boldsymbol{y}$ to the desired STA data $\boldsymbol{x}'$, without any participation of the measurement model $\boldsymbol{H}$. In this way, the network is required to not only describe the geometry of the solution space containing the ground truth STA data $\boldsymbol{x}$, but also the forward mapping $\boldsymbol{H}$. In this study, each matrix element in the PW data $\boldsymbol{y}$ is the linear combination of a whole column of the STA data $\boldsymbol{x}$. However, a convolutional network typically does not have such a large receptive field. Providing the network with an initial solution $\hat{\boldsymbol{x}} = \boldsymbol{H}^T \boldsymbol{y}$ obtained using the conventional LS-STA method could partly relieve this problem, by allowing the network to work as a local filter. However, as the network still solves the inverse problem as a direct mapper, this regression-based method may encounter generalization issues in real applications. Moreover, the deep networks trained in this way typically fail in the preservation of the data consistency between $\boldsymbol{y}' = \boldsymbol{H}\boldsymbol{x}'$ and the measurement $\boldsymbol{y}$. In fact, ensuring data consistency is very important, as it reflects the accuracy of the solution to some degree.

In recent years, model-based learning attracts more and more attention from researchers. These methods combine the conventional model-based iterative methods and the novel data-driven deep learning methods, to achieve accurate and fast solving, respectively. In this way, the forward model $\boldsymbol{H}$ is incorporated into the solving process. Specifically, the iterative method is unrolled to several layers, and the network is trained to replace the gradient or the regularizer parts of the iterative method for acceleration [34][35][36][37].

Null space learning aims at explicitly incorporating the forward model $\boldsymbol{H}$ into the network-based inverse problem solving (to improve the solving accuracy) in an end-to-end manner (to reduce the computational and implementation complexities). In this study, null space learning is introduced in the training of the network to recover the STA dataset $\boldsymbol{x}'$ from the PW dataset $\boldsymbol{y}$.

Given the encoding matrix $\boldsymbol{H} \in \mathbb{R}^{M \times N}$ ($M < N$) and its right pseudo inverse $\boldsymbol{H}^\dagger \in \mathbb{R}^{N \times M}$ ($\boldsymbol{H}\boldsymbol{H}^\dagger = \boldsymbol{I}_M$), it holds that $\mathbb{R}^N = \mathcal{R}(\boldsymbol{H}^\dagger) \oplus \mathcal{N}(\boldsymbol{H})$, where the $\mathcal{R}(\cdot)$ and $\mathcal{N}(\cdot)$ denote the range and null spaces, respectively. That is, for any sample $\boldsymbol{x} \in \mathbb{R}^N$, it can be decomposed as $\boldsymbol{x} = \boldsymbol{x}^+ + \boldsymbol{x}^\perp$, where $\boldsymbol{x}^+ \in \mathcal{R}(\boldsymbol{H}^\dagger)$ and $\boldsymbol{x}^\perp \in \mathcal{N}(\boldsymbol{H})$. Let the operator $P_r \triangleq \boldsymbol{H}^\dagger \boldsymbol{H}$ projects data $\boldsymbol{x}$ to the range space $\mathcal{R}(\boldsymbol{H}^\dagger)$ of $\boldsymbol{H}^\dagger$, and the operator $P_n \triangleq (\boldsymbol{I}_N - \boldsymbol{H}^\dagger(\boldsymbol{H}\boldsymbol{H}^\dagger)\boldsymbol{H})$ projects data $\boldsymbol{x}$ to the null space $\mathcal{N}(\boldsymbol{H})$ of $\boldsymbol{H}$ [38][39][40][41]. A simple illustration of the range-null space decomposition is shown in Fig. 3, and is formulated as follows:

$$x = P_r(x) + P_n(x) \qquad (9)$$

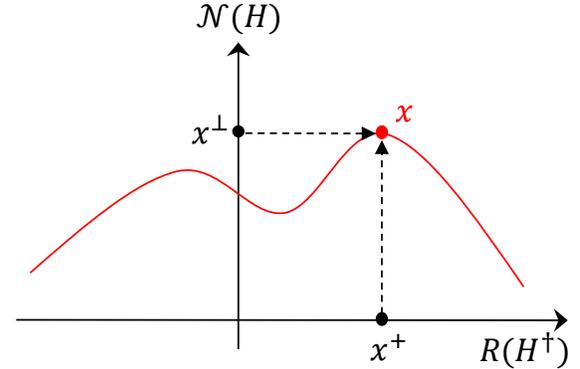

Fig. 3. An illustration of the range-null space decomposition.

For partial Hadamard matrix $\boldsymbol{H}$, its transpose $\boldsymbol{H}^T$ equals its scaled right pseudo inverse $\boldsymbol{H}^\dagger$. Therefore, the solution $\hat{\boldsymbol{x}} = \boldsymbol{H}^T \boldsymbol{y}$ obtained using LS-STA has exact data consistency, as $\boldsymbol{H}\hat{\boldsymbol{x}} \equiv \boldsymbol{H}\boldsymbol{H}^\dagger \boldsymbol{y} \equiv \boldsymbol{y}$. However, by comparing the LS-STA solution $\hat{\boldsymbol{x}} = \boldsymbol{H}^T \boldsymbol{y} = \boldsymbol{H}^T \boldsymbol{H}\boldsymbol{x}$ with Eq. (9), we can find that the LS-STA method neglects the null space component $P_n(\boldsymbol{x})$ of the real STA data $\boldsymbol{x}$, which may be the theoretical explanation for the error introduced in LS-STA-based recovery. In this study, we attempt to estimate the missing null space component with the network to reduce the recovery error.

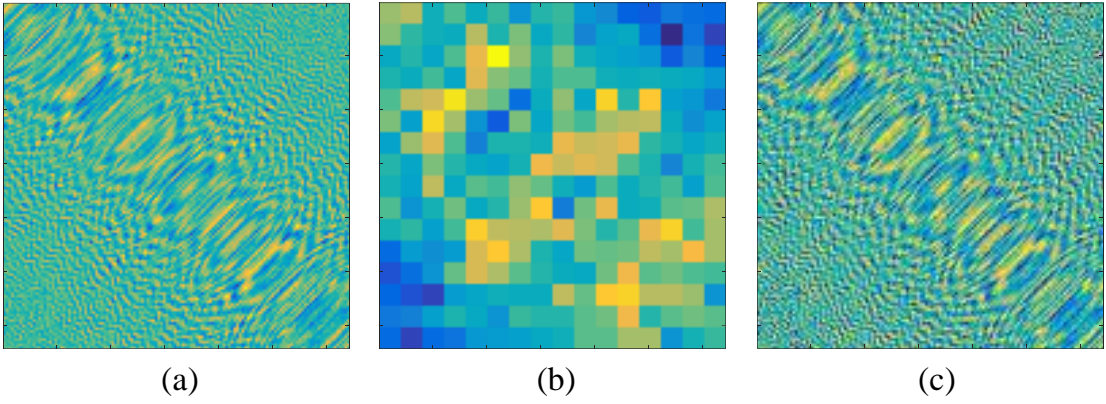

(a)                  (b)                 (c)

Fig. 4. (a) The original ground truth STA data $\boldsymbol{x}$, (b) the local maximum of the small patches, (c) the locally-normalized ground truth STA data $\mathcal{P}(\boldsymbol{x})$.

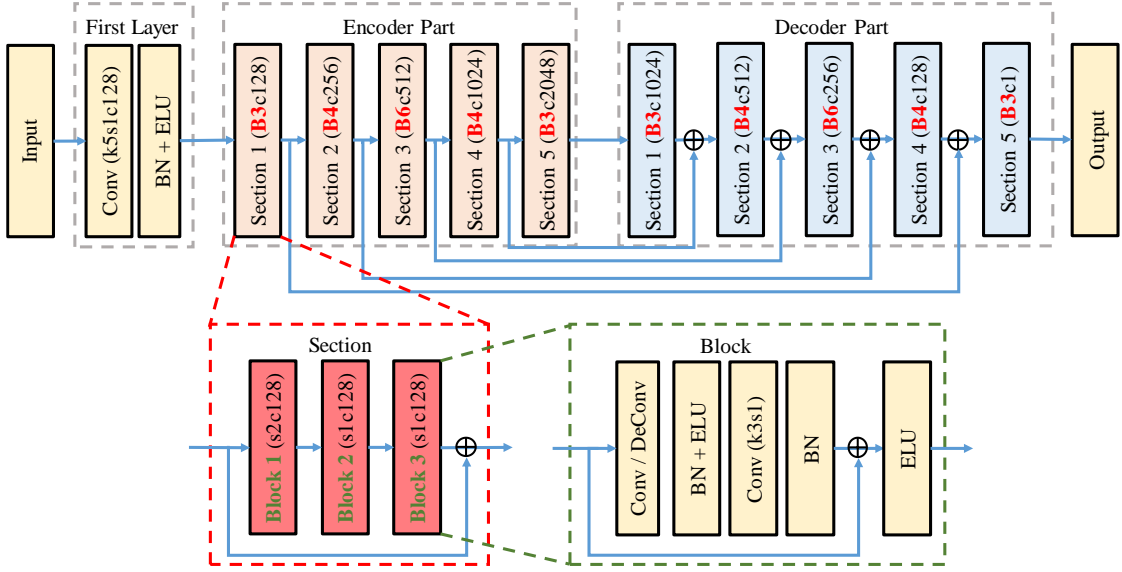

Fig. 5. Illustration of the network architecture. The initial solutions obtained using the LS-STA method are inputted into the network and mapped to the desired output. The kernel size, stride, and the numbers of channels or blocks are indicated by the $k$, $s$, $c$, and $B$, respectively. The blue arrows indicate the skip connections.

Inspired by the null space learning methods [39][40][41], the LS-STA solution is first processed by the trained neural network $\mathcal{F}(\cdot)$ as $\mathcal{F}(\boldsymbol{H}^\dagger \boldsymbol{y})$, and then projected onto the null space of $\boldsymbol{H}$, to produce the missing null space component $P_n(\mathcal{F}(\boldsymbol{H}^\dagger \boldsymbol{y}))$ of the refined STA data. The final refined STA data are:

$$\boldsymbol{x}' = \boldsymbol{H}^\dagger \boldsymbol{y} + P_n(\mathcal{F}(\boldsymbol{H}^\dagger \boldsymbol{y})) \tag{10}$$

which have exact data consistency, as $\boldsymbol{H}\boldsymbol{x}' \equiv \boldsymbol{y}$.

The network is trained by minimizing the mean squared error (MSE) loss between the refined STA data $\boldsymbol{x}'$ and the ground truth STA data $\boldsymbol{x}$. Considering the effects of the element directivity function and the acoustic attenuation, the transmit-receive element pairs with smaller distance would receive echoes with higher intensity. That is, the high intensity signals typically gathers around the diagonal region of the 2D STA data, as shown in Fig. 4(a). Therefore, the STA data are normalized patch-wisely $\mathcal{P}(\cdot)$ before the loss computation, to lead the network focusing not only on the recovery of high-intensity signals but also on the small-intensity signals. Specifically, the STA data are separated into small patches (with a size of $8 \times 8$), which are then normalized to $[-1, 1]$ individually, as shown in Fig. 4.

The complete training loss function is defined as:

$$Loss = \frac{1}{2} \| \mathcal{P}(\boldsymbol{x}') - \mathcal{P}(\boldsymbol{x}) \|_2^2 \tag{11}$$

### C. Neural Network Architecture

The network architecture was designed based on the structures of the U-net convolutional network [42] and the deep residual network (ResNet) [43], as shown in Fig. 5. The first layer takes the input $\hat{\boldsymbol{x}} = \boldsymbol{H}^\dagger \boldsymbol{y}$ obtained using the LS-STA method and increases the number of its feature channels with a relatively large kernel size ($5 \times 5$). The feature maps are subsequently passed through the encoder part of the network, which consists of 5 sections to downsample the data and double the feature channels step by step. Thereafter, the decoder part upsamples the feature maps and halves the feature channels to produce the final output $\mathcal{F}(\boldsymbol{H}^{\dagger}\boldsymbol{y})$. In addition, skip connections are used to pass the features extracted using the encoder sections to the corresponding decoder sections, to recapture the information that might be lost in the downsampling process.

To improve the non-linear mapping capability of the network to enhance its performance, residual blocks [43] are incorporated into the network. The 5 sections in the encoder or decoder parts of the network consist of 3, 4, 6, 4, and 3 residual blocks, respectively. The first block of each section downsamples (by convolutions with a stride of 2) or upsamples (by deconvolutions with a stride of 2) the feature maps, while the following blocks consist of 2 consecutive convolutional layers with strides of 1. In the network, each convolutional layer is followed by the ELU activation function and batch normalization (BN) [44].

### D. Data Acquisition and Network Training

Phantom and *in vivo* data were acquired using an L10-5 linear-array transducer (Shenzhen JiaRui Co., Shenzhen, China), which was connected to a Vantage 256 system (Verasonics Inc., Kirkland, WA, USA). The parameters of the transducer are listed in Table I. Four frames of data were acquired from the carotid artery of a healthy volunteer in the longitudinal (two frames) and the cross-sectional (two frames) views. Two frames of data were acquired from the liver of a healthy volunteer, and a single frame of data was acquired from the biceps of a healthy volunteer. To further enrich the training dataset, Field_II [45] was used to simulate noise-free data from two numerical phantoms (with a size of $60 \times 40 \times 5$ mm$^3$, axial $\times$ lateral $\times$ elevational), both of which contained 4 randomly-positioned hyperechoic cylinders and anechoic cylinders (with radii of 1, 1.5, 2.5 and 3 mm, respectively) and 5 randomly-positioned wire targets. The averaging scattering amplitudes of the hyperechoic cylinders and the wire targets were set to 10 and 100 times higher than that of the background. In the simulations, an acoustic attenuation coefficient of 0.5 dB/cm/MHz was set. The use of the noise-free simulation data can make the network focus on discovering the hidden features without being interfered by noise.

For each acquisition in the phantom and *in-vivo* experiments, 128 STA transmissions and 128 Hadamard-encoded (HE) PW transmissions were performed sequentially with a driving voltage of 15 V to acquire the original complete STA dataset and the full HE-PW dataset, respectively. Thereafter, the received echo signals corresponding to the first 16, 32, and 64 PW transmissions were extracted as partial HE-PW datasets, which would be used to recover the complete STA dataset. The corresponding frame rates were increased by 8-, 4- and 2-fold, respectively. Thanks to the orthogonality of the full Hadamard matrix, the complete STA datasets can be accurately recovered from the full HE-PW transmissions, and can be utilized as the ground truth $x$ for the network training. For each acquisition, 2,300 samples were originally acquired for an imaging depth of 70 mm. Thereafter, each dataset was truncated individually to avoid involving data that only contained noise. For example, there were no meaningful targets or tissue structures located deeper than 45 mm when imaging the carotid artery, and therefore, only the first 1500 samples in the depth direction were preserved as the training data. Eventually, a total number of 14,622 samples were generated for training.

TABLE I
TRANSDUCER PARAMETERS

| Parameter | Value | Unit |
|---|---|---|
| Type | L10-5 | - |
| No. of elements | 128 | - |
| Pitch | 0.3 | mm |
| Kerf | 0.03 | mm |
| Center frequency | 6.25 | MHz |
| Sampling frequency | 25 | MHz |
| Fractional bandwidth | 60% | - |

To evaluate the performance of the proposed method, we acquired testing data from two typical views (wire region and anechoic-target region) of a 040GSE tissue-mimicking phantom (CIRS, Norfolk, VA, USA), the carotid artery in longitudinal and cross-sectional views and the liver of another healthy volunteer.

In this work, three networks were trained to recover the STA dataset from 16, 32, and 64 PW transmissions, and were denoted as CNN-STA-16, CNN-STA-32, and CNN-STA-64, respectively. Because all these networks receive the LS-STA solution (with the size of 128×128) as the input, they share the same architecture. The networks were implemented in Pytorch 1.11.0 [46] on a system with an Intel Xeon Gold 6226R processor and an RTX A6000 GPU (48 GB graphics memory). Each network was trained for 300 epochs with a total training time of ~15 hours. The loss was optimized using the Adam optimizer with a batch size of 64. The initial learning rate was set to $5\times10^{-4}$ and was decayed by 0.95 every 10 epochs.

### E. Evaluation Metrics

To quantitatively evaluate the recovery error of different methods, the normalized root-mean-square error (NRMSE) between the pre-beamformed ground truth STA dataset $x$ (obtained using HE-STA) and the recovered STA dataset $\hat{x}$ (obtained using LS-STA) was calculated as:

$$\text{NRMSE} = \frac{\sqrt{\frac{1}{N^e N^r N^s}\sum_{i=1}^{N^e}\sum_{j=1}^{N^r}\sum_{s=1}^{N^s}(x(e_i,r_j,s_k)-\hat{x}(e_i,r_j,s_k))^2}}{\max\limits_{i,j,k}|x(e_i,r_j,s_k)|} \quad (12)$$

where $N^e$, $N^r$, and $N^s$ denote the total numbers of transmit elements, receive elements, and received samples, respectively. $e_i$, $r_j$ and $s_k$ denote the indexes of transmit element, receive element, and the received sample, respectively. The NRMSE between the pre-beamformed ground truth STA dataset $x$ (obtained using HE-STA) and the refined STA dataset $x'$ (obtained using CNN-STA) can be calculated in a similar way to Eq. (12).

To quantitatively compare the image quality of the LS-STA and CNN-STA methods with different numbers of transmissions (16, 32, and 64), the lateral resolution and contrast were evaluated using full width at half maximum (FWHM) and generalized contrast-to-noise ratio (gCNR) [47], respectively. The STA and HE-STA methods with 128 transmissions were quantitatively evaluated as the baseline.

## IV. RESULTS

### A. Phantom Experiments

Fig. 6 presents the B-mode images for the cyst region of the phantom reconstructed using different methods. Table II presents the computed NRMSEs and gCNRs of targets A, B, and C reconstructed using the CNN-STA and LS-STA methods. The CNN-STA method achieves higher recovery accuracy (lower NRMSEs) than the LS-STA method with the same number of PW transmissions. In addition, the NRMSE decreases as the number of PW transmissions increases for both CNN-STA and LS-STA methods. Owing to the grating lobe artifacts caused by the used λ-pitch array, the gCNR of the anechoic target in the shallow region (ROI A) is lower (with differences of ~ 0.05-0.1) than that of the anechoic target in the deep region (ROI C) in each image. The gCNRs of the anechoic targets (ROIs A and C) increase with the number of PW transmissions, while the gCNRs of the hypoechoic targets (ROI B) are not sensitive to the number of PW transmissions. As for the comparison between the LS-STA and CNN-STA methods, CNN-STA achieves higher or the same gCNRs as LS-STA for almost all the targets (A, B, and C) and all the numbers of transmissions (16, 32, and 64), except for target C with 16 transmissions. The images obtained using the conventional STA and HE-STA methods were quantitatively evaluated for comparison. As target A is located in the shallow region, from which STA imaging could acquire echoes with high-enough SNR [a channel SNR (cSNR) [48] of 23.1 dB was measured in depth from 5 mm to 22 mm (containing target A) with 10 repetitive acquisitions], even though the cSNR can further be improved to 36.7 dB with 128 partial-Hadamard-encoded PW transmissions, this improvement cannot be reflected in the gCNR. Therefore, the STA and HE-STA methods achieve the same gCNR, which is higher than those obtained using LS-STA with 16, 32, and 64 transmissions and CNN-STA with 16 and 32 transmissions. As target C is located in the deep region, from which the STA acquisition would suffer from severe noise [a cSNR of -0.40 dB was measured in depth from 35 mm to 54 mm (containing target C) with 10 repetitive acquisitions], the corresponding gCNR is lower than those of the LS-STA method with 128 transmissions (cSNR: 19.7 dB). These findings are in agreement with the simulation results in [13], that the gCNR of STA image improves when its channel SNR is increased from 0 dB to 10 dB, while the gCNR of STA image remains no change when its channel SNR is further increased from 10 dB to 15 dB.

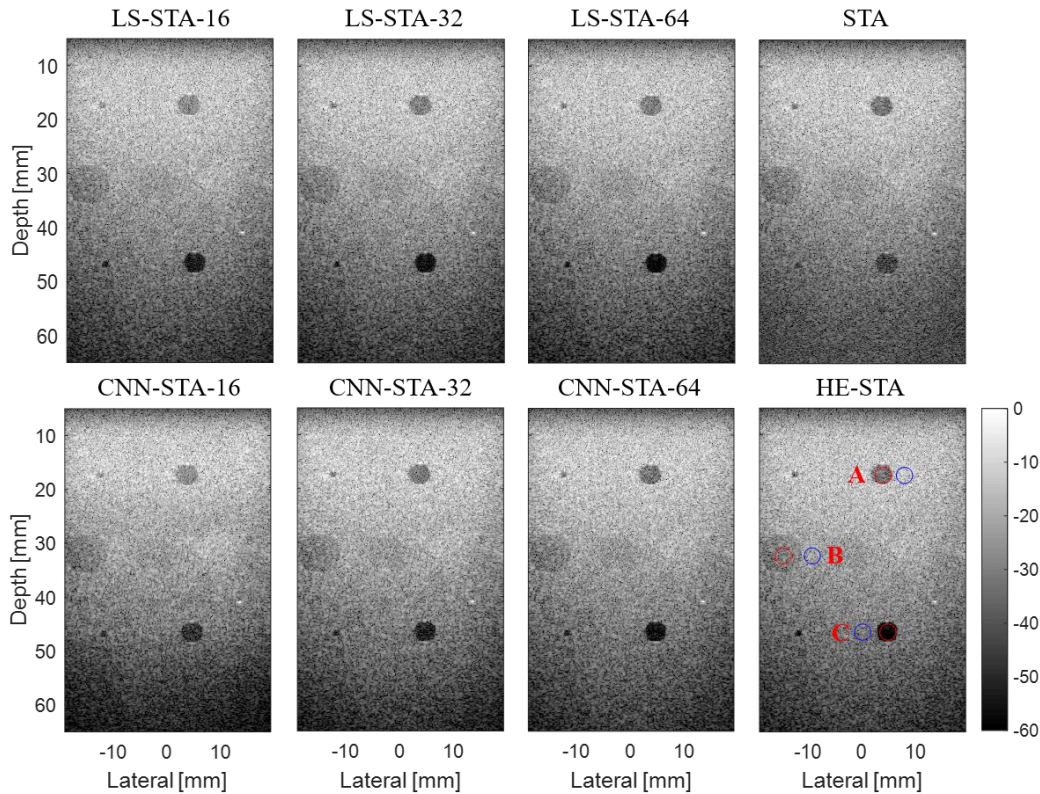

Fig. 6. B-mode images for the cyst region of the phantom reconstructed using the LS-STA and CNN-STA methods with different numbers of transmissions (16, 32, and 64), and the STA and HE-STA methods with 128 transmissions. The red and blue circles indicate the ROIs and background regions used for the calculation of gCNRs.

TABLE II
THE NRMSEs AND gCNRs FOR DIFFERENT METHODS. NOTE THAT LS-STA-128 IS EQUIVALENT TO HE-STA

| Methods | NRMSE [%] | | | gCNR | | | | | | | | | | | |
| --- | --- | --- | --- | --- | --- | --- | --- | --- | --- | --- | --- | --- | --- | --- | --- |
| | 16 Tx | 32 Tx | 64 Tx | 16 Tx | | | 32 Tx | | | 64 Tx | | | 128 Tx | | |
| | | | | A | B | C | A | B | C | A | B | C | A | B | C |
| STA | -- | -- | -- | -- | -- | -- | -- | -- | -- | -- | -- | -- | 0.90 | 0.66 | 0.79 |
| LS-STA | 5.77 | 3.73 | 2.14 | 0.77 | 0.65 | **0.94** | 0.85 | 0.68 | 0.95 | 0.85 | 0.65 | 0.97 | 0.90 | 0.65 | 0.97 |
| CNN-STA | **1.76** | **1.14** | **0.72** | **0.81** | **0.68** | 0.86 | **0.87** | **0.70** | **0.95** | **0.91** | **0.65** | **0.97** | -- | -- | -- |

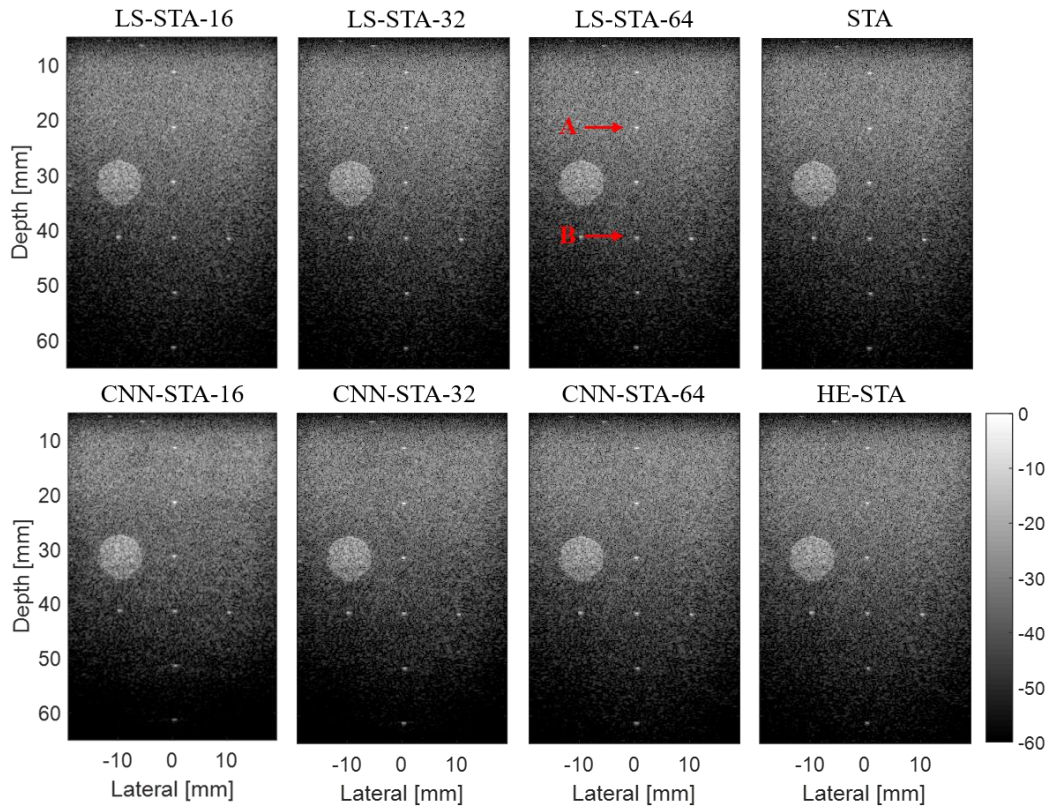

Fig. 7. B-mode images for the wire region of the phantom reconstructed using the LS-STA and CNN-STA methods with different numbers of transmissions (16, 32, and 64), and the STA and HE-STA methods with 128 transmissions. The red arrows indicate the wire targets used for the measurements of the FWHMs.

TABLE III
THE NRMSES AND FWHMS FOR DIFFERENT METHODS. NOTE THAT LS-STA-128 IS EQUIVALENT TO HE-STA

| Methods | NRMSE [%] | | | FWHM [mm] | | | | | | | |
|---|---|---|---|---|---|---|---|---|---|---|---|
| | 16 Tx | 32 Tx | 64 Tx | 16 Tx | | 32 Tx | | 64 Tx | | 128 Tx | |
| | | | | A | B | A | B | A | B | A | B |
| STA | -- | -- | -- | -- | -- | -- | -- | -- | -- | 0.44 | 0.41 |
| LS-STA | 5.50 | 4.42 | 2.58 | 0.44 | 0.41 | 0.44 | 0.41 | 0.44 | 0.41 | 0.44 | 0.41 |
| CNN-STA | **1.90** | **1.36** | **0.85** | 0.44 | 0.42 | 0.42 | 0.41 | 0.45 | 0.42 | -- | -- |

Fig. 7 presents the B-mode images for the wire region of the phantom reconstructed using different methods. The red arrows indicate the wire targets selected for the lateral resolution evaluation. Table III presents the calculated NRMSEs. As shown, the CNN-STA method achieves higher recovery accuracy (lower NRMSE) than LS-STA, and the NRMSEs of the LS-STA and CNN-STA methods decrease as the number of transmissions increases. The computed FWHMs demonstrate that STA, HE-STA, LS-STA, and CNN-STA with all the numbers of transmissions have a very close lateral resolution (with variations of 0.01 mm).

### B. In vivo Experiments

Fig. 8 presents the B-mode images of the cross-sectional carotid artery reconstructed using different methods. As indicated by the blue arrow, severe artifacts exist inside the carotid lumen reconstructed using the LS-STA method. Even though the artifacts can be gradually suppressed as the number of PW transmissions increases, LS-STA-64 still cannot present a clear carotid lumen. In contrast, no obvious artifacts can be observed inside the carotid lumen reconstructed using CNN-STA-16. A red dotted circle and a green dotted circle were manually selected as the ROI and background for the calculation of gCNR. Table IV presents the calculated NRMSEs. As shown, the CNN-STA method achieves higher recovery accuracy (lower NRMSE) than LS-STA, and the NRMSEs of the LS-STA and CNN-STA methods decrease as the number of transmissions increases. In agreement with the results of anechoic target A of the phantom, STA and HE-STA achieve very similar gCNRs, which are higher than those obtained using LS-STA/CNN-STA with all the numbers of transmissions. Note that the proposed CNN-STA method achieves a much higher gCNR than the LS-STA method with the same number of transmissions.

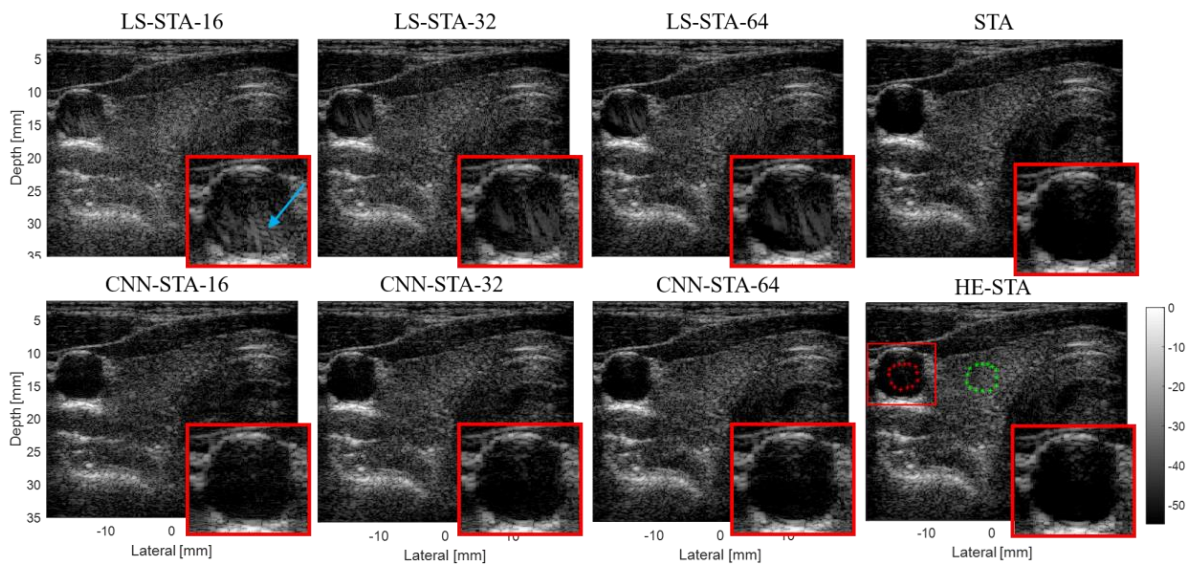

Fig. 8. B-mode images of the cross-sectional carotid artery reconstructed using the LS-STA and CNN-STA methods with different numbers of transmissions (16, 32, and 64), and STA and HE-STA with 128 transmissions. The blue arrow indicates artifacts inside the carotid lumen. The red box on the bottom right of each image is the zoomed-in version of the selected region indicated by the red box in the original image. The red and green dotted circles indicate the manually selected ROI and background region for the gCNR calculation.

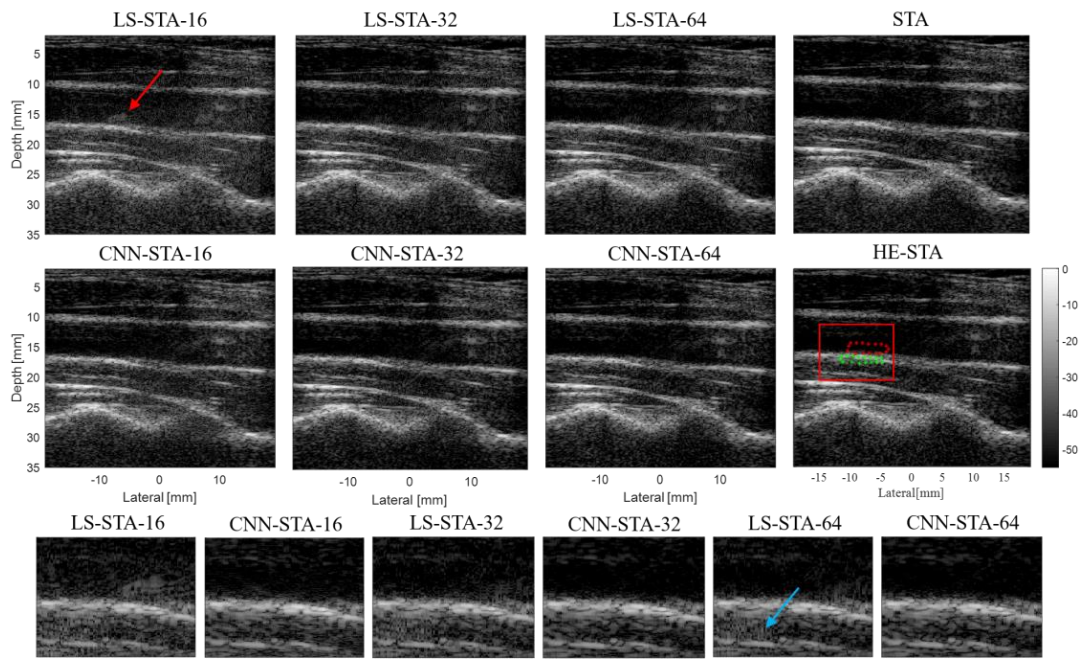

Fig. 9. B-mode images of the longitudinal carotid artery reconstructed using the LS-STA and CNN-STA methods with different numbers of transmissions (16, 32, and 64), STA, and HE-STA with 128 transmissions. The red arrow indicates artifacts inside the carotid lumen. The bottom row images are the zoomed-in version of the selected region indicated by the red box in the original image. The red and green dotted circles indicate the manually selected ROI and background region for the gCNR calculation. The blue arrow indicates strip-like artifacts in the images reconstructed using the LS-STA method.

Fig. 9 presents the B-mode images of the longitudinal carotid

TABLE IV
THE NRMSEs AND gCNRs (OF CROSS-SECTIONAL CAROTID ARTERY) FOR
DIFFERENT METHODS.

| Methods | NRMSE [%] | | | gCNR | | | |
|---|---|---|---|---|---|---|---|
| | 16 Tx | 32 Tx | 64 Tx | 16 Tx | 32 Tx | 64 Tx | 128 Tx |
| STA | -- | -- | -- | -- | -- | -- | 0.85 |
| LS-STA | 2.47 | 1.65 | 0.95 | 0.34 | 0.48 | 0.49 | 0.84 |
| CNN-STA | **0.75** | **0.63** | **0.61** | **0.74** | **0.81** | **0.84** | -- |

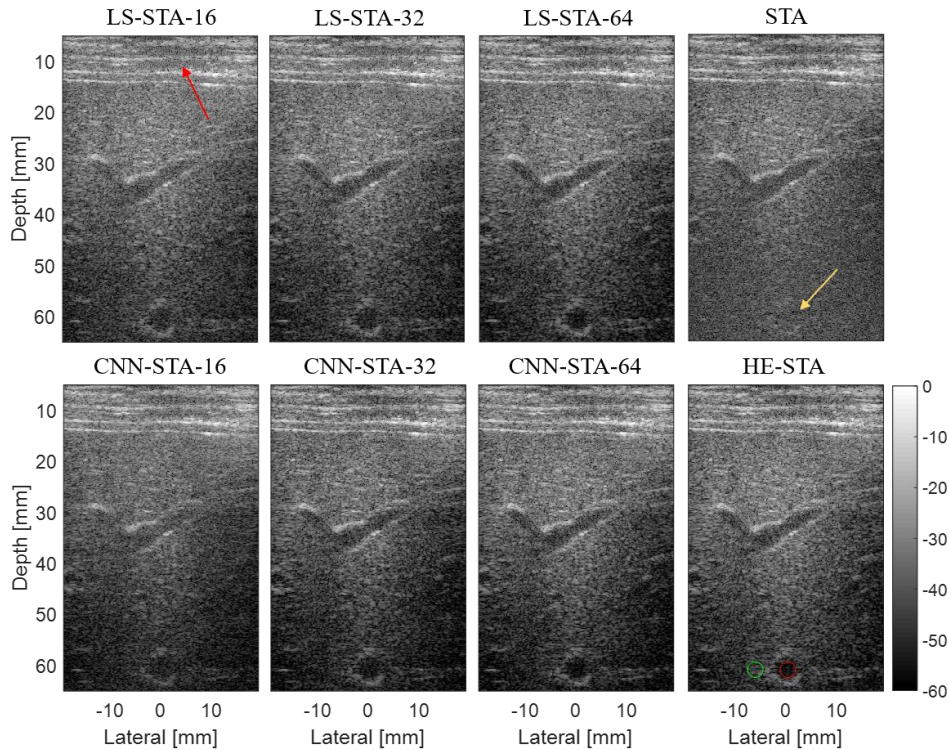

Fig. 10. B-mode images of the liver reconstructed using the LS-STA and CNN-STA methods with different numbers of transmissions (16, 32, and 64), STA, and HE-STA with 128 transmissions. The red and green circles indicate the ROI and the background selected for the gCNR computation. The yellow arrow indicates a vessel lumen which is difficult to discriminate from the background noise in the STA image, but is visible in the LS-STA and CNN-STA images. The red arrow indicates the strip-like artifacts in the shallow region of the image reconstructed using the LS-STA method.

<table>
<tr><th colspan="8">TABLE V<br>THE NRMSEs AND GCNRs (OF LONGITUDINAL CAROTID ARTERY) FOR<br>DIFFERENT METHODS.</th></tr>
</table>

| Methods | NRMSE [%] | | | gCNR | | | |
|---|---|---|---|---|---|---|---|
| | 16 Tx | 32 Tx | 64 Tx | 16 Tx | 32 Tx | 64 Tx | 128 Tx |
| STA | -- | -- | -- | -- | -- | -- | 0.95 |
| LS-STA | 1.59 | 1.03 | 0.59 | 0.83 | 0.91 | 0.93 | 0.95 |
| CNN-STA | **0.47** | **0.32** | **0.28** | **0.92** | **0.95** | **0.95** | -- |

<table>
<tr><th colspan="8">TABLE VI<br>THE NRMSEs AND GCNRs (OF LIVER) FOR DIFFERENT METHODS. NOTE<br>THAT LS-STA-128 IS EQUIVALENT TO HE-STA</th></tr>
</table>

| Methods | NRMSE [%] | | | gCNR | | | |
|---|---|---|---|---|---|---|---|
| | 16 Tx | 32 Tx | 64 Tx | 16 Tx | 32 Tx | 64 Tx | 128 Tx |
| STA | -- | -- | -- | -- | -- | -- | 0.16 |
| LS-STA | 2.55 | 1.69 | 0.98 | **0.32** | 0.46 | 0.59 | 0.66 |
| CNN-STA | **0.74** | **0..59** | **0.51** | 0.26 | **0.47** | **0.63** | -- |

artery reconstructed using different methods. As shown in Table V, the CNN-STA method achieves higher recovery accuracy (lower NRMSE) than the LS-STA method. As a result, CNN-STA can reconstruct images with higher quality. The red arrow indicates artifacts inside the carotid lumen reconstructed using the LS-STA method. These artifacts can be partially suppressed by increasing the number of transmissions. In contrast, the CNN-STA method can present a clear carotid lumen with only 16 transmissions. The images on the bottom row are the zoomed-in version of the selected region indicated by the red box. As indicated by the blue arrow, the LS-STA method introduces strip-like artifacts to the speckles, while the CNN-STA method can recover more realistic speckles. The red and green dotted regions were manually selected as the ROI and background region for the calculation of gCNRs. Results demonstrate that CNN-STA performs better than the LS-STA method with the same number of transmissions, and the gCNRs of both methods increase with the number of transmissions. The CNN-STA method can achieve a similar gCNR to the STA and HE-STA methods with only 32 transmissions.

Fig. 10 presents the B-mode images of the liver reconstructed using different methods. As shown, the contrast of the image reconstructed using the STA method is significantly lower than those of the other methods. For example, it is difficult to discriminate the vessel lumen indicated by the yellow arrow with the background noise in the STA image. The strip-like artifacts can also be observed in the shallow region of the images reconstructed using the LS-STA method, as indicated by the red arrow. The red and green circles were selected as the ROI and background region for the gCNR calculation. As shown in Table VI, the gCNRs of the LS-STA and CNN-STA methods increase with the number of transmissions, and both methods with only 16 transmissions can achieve higher gCNR than STA with 128 transmissions. As for the comparison between the LS-STA and CNN-STA methods, CNN-STA performs worse than LS-STA with 16 transmissions, but better than LS-STA with 32 and 64 transmissions.

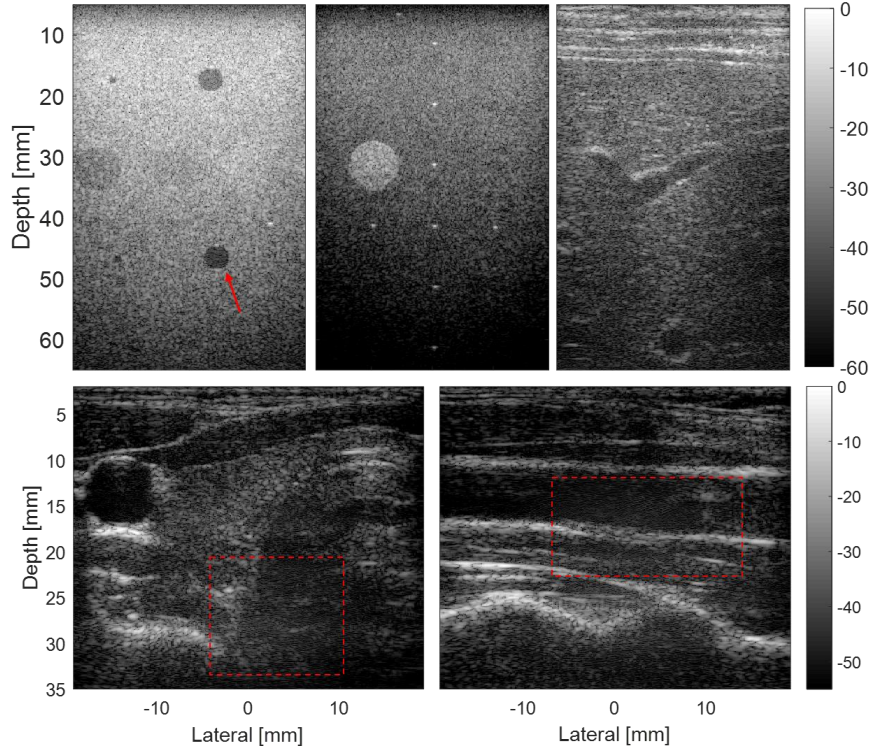

Fig. 11. B-mode images of the phantom, liver, and carotid artery reconstructed using pCNN-STA-32 trained without null space learning. Compared with the images reconstructed using CNN-STA-32, a decrease in the contrast of the anechoic target indicated by the red arrow can be observed. In addition, more clutters are present in the reconstructed images.

### C. Ablation Experiments

To demonstrate the effectiveness of the null space learning method, the output of the network $\mathcal{F}(\boldsymbol{H}^{\dagger}\boldsymbol{y})$ was no longer projected onto the null space of $\boldsymbol{H}$ and was directly taken as the final refined STA data. Correspondingly, the loss function was changed to the MSE between the network output $\mathcal{F}(\boldsymbol{H}^{\dagger}\boldsymbol{y})$ and the ground truth STA data $x$.

The network trained using the above loss function was named plain CNN-STA to recover the complete STA dataset from 32 PW transmissions (pCNN-STA-32). The reconstructed phantom, liver, and carotid images are demonstrated in Fig. 11 for comparison. As indicated by the red arrow, more noise can be observed in the anechoic cyst than in the CNN-STA-32 image (Fig. 6). In addition, clutters can be observed in the background, as indicated by the red dotted boxes. The gCNRs and FWHMs were calculated for quantitative comparison. The gCNRs of targets A, B, and C in the cyst region of the phantom reconstructed using pCNN-STA-32 are 0.86, 0.65, and 0.9, respectively, which are all lower than those obtained using CNN-STA-32 (0.87, 0.70, and 0.95). The FWHMs of wire targets A and B in the wire region of the phantom reconstructed using pCNN-STA-32 are 0.42 and 0.42, respectively, which are comparable to or slightly worse than those obtained using

CNN-STA-32 (0.42 and 0.41). The gCNRs of the selected targets in the liver and carotid artery in the cross-sectional and longitudinal views are 0.35, 0.79, and 0.93, respectively, which are lower than those obtained using CNN-STA-32 (0.47, 0.81, and 0.95).

### D. Computational Complexity and Speed

The number of floating-point operations (FLOPs) required for CNN-STA ($4.2 \times 10^8$ parameters) is 65.8 billion, and the used NVIDIA RTX A6000 GPU can process 38.7 TFLOPs per second. Theoretically, CNN-STA can process 595 samples per second. In this study, the complete STA dataset has 2,300 samples in the axial direction, which results in a total computational time of 3.87 s. In real experiments, the measured recovery time is about 7.14 s (averaged from 50 draws), owing to the extra data transfer time and additional null-space mapping time.

### V. DISCUSSIONS

The results demonstrate that the proposed CNN-STA method can effectively reduce the error in the STA dataset recovered using the LS-STA method, by estimating its missing null space component. As a result, the proposed method typically reconstructs images with higher quality (both qualitatively and quantitatively) than the LS-STA method, especially in the shallow region. For example, the CNN-STA method presents a very clear carotid lumen with only 16 PW transmissions, while LS-STA suffers from high-level artifacts even with 64 PW transmissions, as shown in Figs. 8 and 9. The CNN-STA method achieves higher gCNRs than the LS-STA method for almost all the targets and all the numbers of transmissions, except for target C with 16 transmissions. The same phenomenon can be observed in liver images with 16 transmissions, where the gCNR of the vessel lumen located in the deep region reconstructed using CNN-STA is lower than that reconstructed using LS-STA. When the number of transmissions is increased to 32 or 64, the CNN-STA method can achieve better performance than the LS-STA method. This is probably because, in situations where the degree of underdetermination is high (i.e., the measurements are limited), the noise in echoes backscattered from the deep region would more severely interfere with the feature extraction of the network. The proposed CNN-STA method can maintain the high resolution of STA imaging with a small number of transmissions (as low as 16), as LS-STA does. The NRMSE index describes the recovery accuracy of the STA dataset. Typically, a lower NRMSE corresponds to better image quality, including the quantitative metrics. However, in the cyst phantom and liver images with 16 transmissions, the NRMSEs of the CNN-STA method is lower than those of the LS-STA method, while the gCNRs (for target C and the vessel lumen) of the CNN-STA method are lower than those of the LS-STA method. A possible explanation for these results is as follows. The NRMSE describes the overall consistency between the compared STA datasets. Owing to the acoustic attenuation, the intensities of the samples in the STA dataset have a decrease trend along the depth direction. Therefore, even though CNN-STA has a higher recovery error in the deep region than LS-STA with 16 transmissions, its contribution to the NRMSE calculation is relatively small.

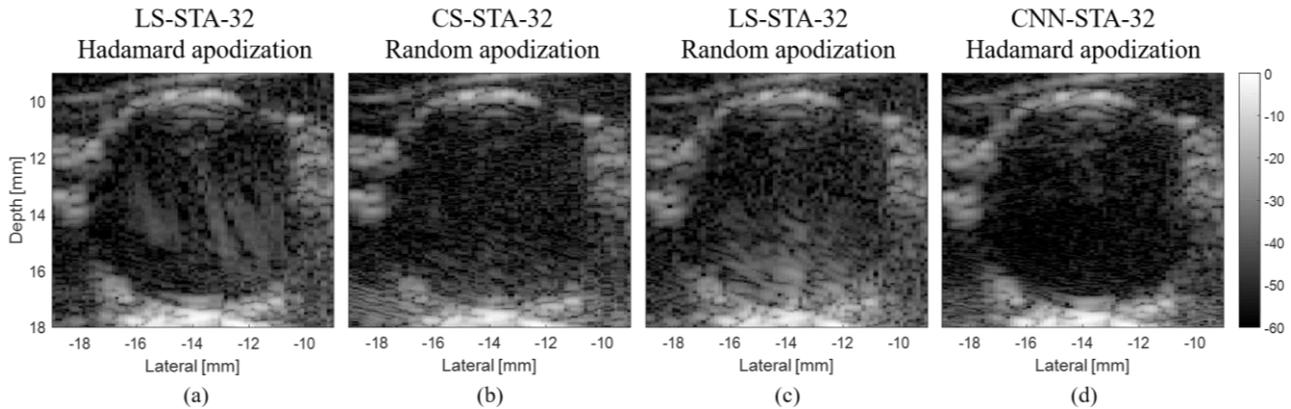

Fig. 12. B-mode images of the carotid lumen reconstructed using (a) the LS-STA method with 32 partial-Hadamard-encoded PW transmissions, (b) the CS-STA method with 32 randomly encoded PW transmissions, (c) the LS-STA method (using Tikhonov regularization) with 32 randomly encoded PW transmissions and (d) the proposed CNN-STA method with 32 partial-Hadamard-encoded PW transmissions.

It should be noted that the forms of the artifacts in the reconstructed images are related to the types of recovery algorithms and measurement matrices. In the early study of CS-STA, a uniformly random matrix was used to encode the PW transmissions in the linear array configuration [9]. As demonstrated in Fig. 12(b) and [9], if the PW transmissions are encoded with random apodizations (i.e, $H_{m,n}$ obeys a continuous uniform distribution) and the STA dataset is recovered using the CS-STA method, the block-like artifacts in the image [Fig. 12(a)] (reconstructed from partial Hadamard encoded PW transmissions using the LS-STA method) are spread to the whole lumen as many isolated small artifacts [Fig. 12(b)]. In this way, the tissue structures can be unobstructedly presented, and the gCNR of the carotid lumen is improved to 0.64, which is higher than that of LS-STA with Hadamard-encoded apodizations (0.48). However, its gCNR is lower than those of STA (0.77) and CNN-STA (0.81) with the same number of transmissions (32). Therefore, using the proposed method to refine the recovered STA dataset is still necessary. Note that when the PW transmissions are encoded with a random matrix, the CS-STA method is very time-consuming (~1 hour with parallel computing) and thus is not appropriate for generating the initial solution for the proposed CNN-STA method. If we would like to achieve fast solving with the LS-STA method, considering that the partial random matrix $H$ is not invertible, we need to introduce Tikhonov regularization [49][50] to ensure the robustness of the solution. As shown in Fig. 12(c), severe artifacts could still be observed in the bottom of the vessel lumen, and the measured gCNR is only 0.52. In summary, with the current encoding matrices (random/Hadamard) and the reconstruction algorithms (CS/LS-STA), we have to find a trade-off between the reconstruction speed and image quality. Considering the transmitted energy (Hadamard encoding can increase the average transmitted energy by 2-fold, compared with random encoding) and the computational efficiency in obtaining the initial solution for the network [14], we used the Hadamard encoding matrix (instead of random encoding matrix) and the LS-STA method (instead of CS-STA) in the proposed method. Recently, Chen et al proposed a deep learning method named ApodNet [51] to optimize the transmit apodizations (encoding matrix). As the encoding matrix affects the quality of the initial solution obtained using the CS/LS-STA methods, combination of ApodNet and the proposed CNN-STA method may further improve the accuracy of STA dataset recovery, and remains for future study.

The labels used for the training of the network were obtained using the HE-STA method. Therefore, the trained network cannot perform better than HE-STA imaging, qualitatively or quantitatively. However, we should note that the proposed method achieves close results to the HE-STA method with a significantly higher frame rate (up to 8 times). Phantom and liver experiments demonstrate that the proposed CNN-STA method can achieve much higher gCNRs than STA imaging (with 128 transmissions) in the deep region. The carotid experiments demonstrate that the performance of CNN-STA is worse than that of STA imaging (with 128 transmissions) in the shallow region. However, we should note that the frame rate of the proposed method is much higher than that of the STA imaging (by up to 8 times), only with a small sacrifice of the image quality in the shallow region.

In this study, the 2D STA data in the Rx-Tx space was input to the CNN-STA to exploit its symmetrical property, ripple-like structures, and other hidden features to improve its recovery performance. There are more meaningful properties in the fast-time dimension of ultrasound signals to exploit, including its limited bandwidth and the time-of-flight relationship. Therefore, feeding the 3D STA data (samples-Rx-Tx) to the network may further enhance the STA dataset recovery performance. Expanding the dimension of the input data will undoubtedly increase the computational complexity and time. More experiments should be carried out to find an appropriate trade-off between these factors. In this study, a very deep network was designed to achieve higher recovery accuracy. We will investigate whether a smaller network can achieve the same or close performance to improve the computational efficiency for real applications. The training dataset includes data acquired from the carotid artery and the liver with relatively mild motion. Therefore, the trained network is deemed to have some level of robustness to motion artifacts. However, it is necessary to investigate the performance of the network when the object moves fast.

Compared with other image-domain end-to-end networks, which directly map the low-quality images to the high-quality images, the proposed CNN-STA method has to carry out the non-linear mapping thousands of times to recover the complete STA dataset and obtain the corresponding high-quality images. There is no doubt that the proposed method has much higher computational complexity and is more time-consuming than the image-domain methods. However, by outputting the raw RF data rather than the B-mode image, the proposed method may enable subsequent shear wave elastography [52] and vector flow imaging [53]. In addition, these novel imaging modes may benefit from the high-frame-rate, high-resolution, and high-penetration performances of the proposed CNN-STA method.

Currently, the inference frame rate of the proposed method is far lower than the requirements for real-time imaging. The inference speed can be improved with several means as follows. First, the trained network could be deployed in NVIDIA TensorRT (SDK for high-performance deep learning inference) without any modifications to its architecture for acceleration. Second, we could reduce the number of layers or the size of the convolutional kernels of the network to find a good trade-off between the computational time and the recovery accuracy.

In this study, the STA channel datasets recovered using LS-STA or refined using CNN-STA were beamformed using the conventional DAS method. It is promising to apply adaptive or other network-based beamforming methods to further improve the final image quality.

## VI. Conclusion

In this work, we train a convolutional neural network to refine the complete STA dataset recovered from partial Hadamard-encoded PW transmissions using our previously proposed LS-STA method, by estimating the missing null space component of the STA data. Results demonstrate that the proposed method can effectively reduce the recovery error of LS-STA to suppress the artifacts, especially in the shallow region. In addition, the proposed method maintains the high resolution of STA imaging and achieves higher contrast in the deep region than STA imaging with fewer PW transmissions, as LS-STA does. These advantages pave the way for the proposed method to real applications.

## References

[1] J. A. Jensen, S. I. Nikolov, K. L. Gammelmark, and M. H. Pedersen, "Synthetic aperture ultrasound imaging," *Ultrasonics*, vol. 44, no. SUPPL., pp. e5–e15, Dec. 2006.

[2] M. Karaman, P. C. Li, and M. O'Donnell, "Synthetic aperture imaging for small scale systems," *IEEE Trans. Ultrason. Ferroelectr. Freq. Control*, vol. 42, no. 3, pp. 429–442, 1995.

[3] M. H. Bae, "A study of synthetic-aperture imaging with virtual source elements in B-mode ultrasound imaging systems," *IEEE Trans. Ultrason. Ferroelectr. Freq. Control*, vol. 47, no. 6, pp. 1510–1519, 2000.

[4] R. Y. Chiao, L. J. Thomas, and S. D. Silverstein, "Sparse array imaging with spatially-encoded transmits," *Proc. IEEE Ultrason. Symp.*, vol. 2, no. c, pp. 1679–1682, 1997.

[5] T. Harrison, A. Sampaleanu, and R. Zemp, "S-sequence spatially-encoded synthetic aperture ultrasound imaging [Correspondence]," *IEEE Trans. Ultrason. Ferroelectr. Freq. Control*, vol. 61, no. 5, pp. 886–890, 2014.

[6] P. Gong, M. C. Kolios, and Y. Xu, "Delay-encoded transmission and image reconstruction method in synthetic transmit aperture imaging," *IEEE Trans. Ultrason. Ferroelectr. Freq. Control*, vol. 62, no. 10, pp. 1745–1756, 2015.

[7] M. O'Donnell and Y. Wang, "Coded excitation for synthetic aperture ultrasound imaging," *IEEE Trans. Ultrason. Ferroelectr. Freq. Control*, vol. 52, no. 2, pp. 171–176, 2005.

[8] N. Bottenus, "Recovery of the complete data set from focused transmit beams," *IEEE Trans. Ultrason. Ferroelectr. Freq. Control*, vol. 65, no. 1, pp. 30–38, 2018.

[9] J. Liu, Q. He, and J. Luo, "A compressed sensing strategy for synthetic transmit aperture ultrasound imaging," *IEEE Trans. Med. Imaging*, vol. 36, no. 4, pp. 878–891, 2017.

[10] J. Liu and J. Luo, "Compressed sensing based synthetic transmit aperture for phased array using Hadamard encoded diverging wave transmission," *IEEE Trans. Ultrason. Ferroelectr. Freq. Control*, vol. 65, no. 7, pp. 1141–1152, Jul. 2018.

[11] J. Liu, Q. He, and J. Luo, "Compressed sensing based synthetic transmit aperture imaging: validation in a convex array configuration," *IEEE Trans. Ultrason. Ferroelectr. Freq. Control*, vol. 65, no. 3, pp. 300–315, Mar. 2018.

[12] Y. Chen, J. Liu, J. Grondin, E. E. Konofagou, and J. Luo, "Compressed sensing reconstruction of synthetic transmit aperture dataset for volumetric diverging wave imaging," *Phys. Med. Biol.*, vol. 64, no. 2, p. 025013, Jan. 2019.

[13] J. Zhang *et al.*, "Acceleration of reconstruction for compressed sensing based synthetic transmit aperture imaging by using in-phase/quadrature data," *Ultrasonics*, vol. 118, no. May, p. 106576, 2021.

[14] J. Zhang, J. Liu, W. Fan, W. Qiu, and J. Luo, "Partial Hadamard encoded synthetic transmit aperture for high frame rate imaging with minimal l 2 -norm least squares method," *Phys. Med. Biol.*, vol. 67, no. 10, p. 105002, May 2022.

[15] R. J. G. van Sloun, R. Cohen, and Y. C. Eldar, "Deep learning in ultrasound imaging," *Proc. IEEE*, vol. 108, no. 1, pp. 11–29, Jan. 2020.

[16] M. Gasse, F. Millioz, E. Roux, D. Garcia, H. Liebgott, and D. Friboulet, "High-quality plane wave compounding using convolutional neural

networks," *IEEE Trans. Ultrason. Ferroelectr. Freq. Control*, vol. 64, no. 10, pp. 1637–1639, Oct. 2017.

[17] X. Zhang, J. Li, Q. He, H. Zhang, and J. Luo, "High-quality reconstruction of plane-wave imaging using generative adversarial network," in *2018 IEEE International Ultrasonics Symposium (IUS)*, Oct. 2018, pp. 1–4.

[18] J. Lu, F. Millioz, D. Garcia, S. Salles, D. Ye, and D. Friboulet, "Complex convolutional neural networks for ultrafast ultrasound imaging reconstruction from in-phase/quadrature signal," *IEEE Trans. Ultrason. Ferroelectr. Freq. Control*, vol. 69, no. 2, pp. 592–603, 2022.

[19] Z. Zhou, Y. Wang, Y. Guo, X. Jiang, and Y. Qi, "Ultrafast plane wave imaging with line-scan-quality using an ultrasound-transfer generative adversarial network," *IEEE J. Biomed. Heal. Informatics*, vol. 24, no. 4, pp. 943–956, 2020.

[20] D. Perdios, M. Vonlanthen, F. Martinez, M. Arditi, and J.-P. Thiran, "CNN-based image reconstruction method for ultrafast ultrasound imaging," *IEEE Trans. Ultrason. Ferroelectr. Freq. Control*, vol. 69, no. 4, pp. 1154–1168, Apr. 2022.

[21] D. Perdios, M. Vonlanthen, F. Martinez, M. Arditi, and J. P. Thiran, "CNN-based ultrasound image reconstruction for ultrafast displacement tracking," *IEEE Trans. Med. Imaging*, vol. 40, no. 3, pp. 1078–1089, 2021.

[22] H. Zuo, J. Zhang, J. Luo, and B. Peng, "Phase constraint improves ultrasound image quality reconstructed using deep neural network," *IEEE Int. Ultrason. Symp. IUS*, pp. 1–4, 2021.

[23] J. Zhang, Q. He, Y. Xiao, H. Zheng, C. Wang, and J. Luo, "Ultrasound image reconstruction from plane wave radio-frequency data by self-supervised deep neural network," *Med. Image Anal.*, vol. 70, p. 102018, May 2021.

[24] B. Luijten *et al.*, "Adaptive ultrasound beamforming using deep learning," *IEEE Trans. Med. Imaging*, vol. 39, no. 12, pp. 3967–3978, Dec. 2020.

[25] A. Wiacek, E. Gonzalez, and M. A. L. Bell, "CohereNet: a deep learning architecture for ultrasound spatial correlation estimation and coherence-based beamforming," *IEEE Trans. Ultrason. Ferroelectr. Freq. Control*, vol. 67, no. 12, pp. 2574–2583, Dec. 2020.

[26] S. Khan, J. Huh, and J. C. Ye, "Switchable and tunable deep beamformer using adaptive instance normalization for medical ultrasound," *IEEE Trans. Med. Imaging*, vol. 41, no. 2, pp. 266–278, 2021.

[27] X. Huang, M. A. Lediju Bell, and K. Ding, "Deep learning for ultrasound beamforming in flexible array transducer," *IEEE Trans. Med. Imaging*, vol. 40, no. 11, pp. 3178–3189, 2021.

[28] D. Hyun, L. L. Brickson, K. T. Looby, and J. J. Dahl, "Beamforming and speckle reduction using neural networks," *IEEE Trans. Ultrason. Ferroelectr. Freq. Control*, vol. 66, no. 5, pp. 898–910, 2019.

[29] A. A. Nair, K. N. Washington, T. D. Tran, A. Reiter, and M. A. Lediju Bell, "Deep learning to obtain simultaneous image and segmentation outputs from a single input of raw ultrasound channel data," *IEEE Trans. Ultrason. Ferroelectr. Freq. Control*, vol. 67, no.12, pp. 2493–2509, Dec. 2020.

[30] L. L. Brickson, D. Hyun, M. Jakovljevic, and J. J. Dahl, "Reverberation noise suppression in ultrasound channel signals using a 3D fully convolutional neural network," *IEEE Trans. Med. Imaging*, vol. 40, no. 4, pp. 1184–1195, 2021.

[31] A. C. Luchies and B. C. Byram, "Deep neural networks for ultrasound beamforming," *IEEE Trans. Med. Imaging*, vol. 37, no. 9, pp. 2010–2021, Sep. 2018.

[32] Y. H. Yoon, S. Khan, J. Huh, and J. C. Ye, "Efficient B-mode ultrasound image reconstruction from sub-sampled RF data using deep learning," *IEEE Trans. Med. Imaging*, vol. 38, no. 2, pp. 325–336, 2019.

[33] I. A. M. Huijben, B. S. Veeling, K. Janse, M. Mischi, and R. J. G. Van Sloun, "Learning sub-sampling and signal recovery with applications in ultrasound imaging," *IEEE Trans. Med. Imaging*, vol. 39, no. 12, pp. 3955–3966, 2020.

[34] O. Solomon *et al.*, "Deep unfolded robust PCA with application to clutter suppression in ultrasound," *IEEE Trans. Med. Imaging*, vol. 39, no. 4, pp. 1051–1063, Apr. 2020.

[35] A. Mamistvalov and Y. C. Eldar, "Deep unfolded recovery of sub-Nyquist sampled ultrasound images," *IEEE Trans. Ultrason. Ferroelectr. Freq. Control*, vol. 68, no. 12, pp. 3484–3496, 2021.

[36] H. Zhang, B. Liu, H. Yu, and B. Dong, "MetaInv-Net: meta inversion network for sparse view CT image reconstruction," *IEEE Trans. Med. Imaging*, vol. 40, no. 2, pp. 621–634, 2021.

[37] Y. Yang, J. Sun, H. Li, and Z. Xu, "ADMM-CSNet: A deep learning approach for image compressive sensing," *IEEE Trans. Pattern Anal. Mach. Intell.*, vol. 42, no. 3, pp. 521–538, 2020.

[38] D. Bertsekas, *Nonlinear Programming*. Belmont, MA, 1999.

[39] J. Schwab, S. Antholzer, and M. Haltmeier, "Deep null space learning for inverse problems: convergence analysis and rates," *Inverse Probl.*, vol. 35, no. 2, p. 025008, Feb. 2019.

[40] D. Chen and M. E. Davies, "Deep decomposition learning for inverse imaging problems," in *Lecture Notes in Computer Science (including subseries Lecture Notes in Artificial Intelligence and Lecture Notes in Bioinformatics)*, vol. 12373 LNCS, no. 1, 2020, pp. 510–526.

[41] C. K. Sønderby, J. Caballero, L. Theis, W. Shi, and F. Huszár, "Amortised map inference for image super-resolution," *5th Int. Conf. Learn. Represent. ICLR 2017 - Conf. Track Proc.*, pp. 1–17, 2017.

[42] O. Ronneberger, P. Fischer, and T. Brox, "U-Net: convolutional networks for biomedical image segmentation," *Med. Image Comput. Comput. Assist. Interv.*, pp. 234–241, 2015.

[43] K. He, X. Zhang, S. Ren, and J. Sun, "Deep residual learning for image recognition," *Proc. IEEE Comput. Soc. Conf. Comput. Vis. Pattern Recognit.*, vol. 2016-Decem, pp. 770–778, 2016.

[44] S. Ioffe and C. Szegedy, "Batch normalization : accelerating deep network training by reducing internal covariate shift," *arXiv: 1502.03167v3*, 2015.

[45] J. A. Jensen and N. B. Svendsen, "Calculation of pressure fields from arbitrarily shaped, apodized, and excited ultrasound transducers," *IEEE Trans. Ultrason. Ferroelectr. Freq. Control*, vol. 39, no. 2, pp. 262–267, Mar. 1992.

[46] A. Paszke *et al.*, "Automatic differentiation in Pytorch," 2017. doi://10.1145/24680.24681.

[47] J. Jensen, M. B. Stuart, and J. A. Jensen, "Optimized plane wave imaging for fast and high-quality ultrasound imaging," *IEEE Trans. Ultrason. Ferroelectr. Freq. Control*, vol. 63, no. 11, pp. 1922–1934, 2016.

[48] A. Rodriguez-Molares *et al.*, "The generalized contrast-to-noise ratio: a formal definition for lesion Detectability," *IEEE Trans. Ultrason. Ferroelectr. Freq. Control*, vol. 67, no. 4, pp. 745–759, 2020.

[49] B. Berthon *et al.*, "Spatiotemporal matrix image formation for programmable ultrasound scanners," *Phys. Med. Biol.*, vol. 63, no. 3, p. 03NT03, Feb. 2018.

[50] Q. You, Z. Dong, M. R. Lowerison, and P. Song, "Pixel-oriented adaptive apodization for plane-wave imaging based on recovery of the complete dataset," *IEEE Trans. Ultrason. Ferroelectr. Freq. Control*, vol. 69, no. 2, pp. 512–522, 2022.

[51] Y. Chen, J. Liu, X. Luo, and J. Luo, "ApodNet: Learning for high frame rate synthetic transmit aperture ultrasound imaging," *IEEE Trans. Med. Imaging*, vol. 40, no. 11, pp. 3190–3204, Nov. 2021.

[52] G. Montaldo, M. Tanter, J. Bercoff, N. Benech, and M. Fink, "Coherent plane-wave compounding for very high frame rate ultrasonography and transient elastography," *IEEE Trans. Ultrason. Ferroelectr. Freq. Control*, vol. 56, no. 3, pp. 489–506, Mar. 2009.

[53] J. A. Jensen, S. I. Nikolov, A. C. H. Yu, and D. Garcia, "Ultrasound vector flow imaging-part I: sequential systems," *IEEE Trans. Ultrason. Ferroelectr. Freq. Control*, vol. 63, no. 11, pp. 1704–1721, 2016.
.

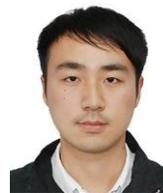
**Jingke Zhang** was born in Hubei, China, in 1996. He received the B.S. degree in Biomedical Engineering from Huazhong University of Science and Technology, Wuhan, China, in 2017. He is currently pursuing the Ph.D. degree with the department of biomedical engineering, Tsinghua University, Beijing, China. His research interests include sparse regularization-based and deep learning-based ultrasound beamforming, compressed sensing and ultrasound localization microscopy.

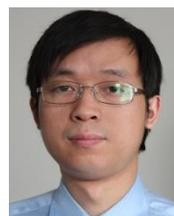
**Jianwen Luo** (S'02–M'06–SM'14) received the B.S. and Ph.D. degrees (Hons.) in biomedical engineering from Tsinghua University, Beijing, China, in 2000 and 2005, respectively. He was a Postdoctoral Research Scientist from 2005 to 2009, and an Associate Research Scientist from 2009 to 2011 with the Department of Biomedical Engineering, Columbia University, New York, NY, USA. He joined the Department of Biomedical Engineering and the Center for Biomedical Imaging Research at Tsinghua University as a Tenure Track Associate Professor, and became a Tenure Associate Professor in 2017.

He was enrolled in the Thousand Young Talents Program of China in 2012, and received the Excellent Young Scientists Fund from the National Natural Science Foundation of China (NSFC) in 2013. He was supported by the National Key R&D Program of China in 2016 and 2020. He received several awards including Chinese Society of Biomedical Engineering (CSMBE) Huang Jiasi Award, China Society of Image and Graphics Technological Invention, Jiangsu Province Science and Technology Award and Beijing Science and Technology Award. He has authored or coauthored over 170 peer-reviewed articles in international journals, 100 conference proceedings papers, and 200 conference abstracts. His research interests include ultrasound imaging, fluorescence imaging, and photoacoustic imaging.

He serves as a member of the IEEE Engineering in Medicine and Biology Society (EMBS) Technical Committee on Biomedical Imaging and Image Processing (BIIP) and the Technical Program Committee of IEEE International Ultrasonics Symposium (IUS). He serves as an Associate Editor for IEEE Transactions on Ultrasonics, Ferroelectrics, And Frequency Control and an Advisory Editorial Board Member of the Journal of Ultrasound in Medicine and an Editorial Board Member of Ultrasonics.